# Thermoplastic deformation of silicon surfaces induced by ultrashort pulsed lasers in submelting conditions

G. D. Tsibidis [1♣], E. Stratakis [1,2], and K. E. Aifantis [3,4]

[1] *Institute of Electronic Structure and Laser (IESL), Foundation for Research and Technology (FORTH), N. Plastira 100, Vassilika Vouton, 70013, Heraklion, Crete, Greece*

[2] *Department of Materials Science and Technology, University of Crete, 710 03 Heraklion, Crete, Greece.*

[3] *Laboratory of Mechanics and Materials, Aristotle University of Thessaloniki, GR-54006 Thessaloniki, Greece*

[4] *Department of Physics, Michigan Technological University, Houghton, MI 49931 USA*

A hybrid theoretical model is presented to describe thermoplastic deformation effects on silicon surfaces induced by single and multiple ultrashort pulsed laser irradiation in submelting conditions. An approximation of the Boltzmann transport equation is adopted to describe the laser irradiation process. The evolution of the induced deformation field is described initially by adopting the differential equations of dynamic thermoelasticity while the onset of plastic yielding is described by the von Mise's stress. Details of the resulting picometre sized crater, produced by irradiation with a single pulse, are discussed as a function of the imposed conditions and thresholds for the onset of plasticity are computed. Irradiation with multiple pulses leads to ripple formation of nanometre size that originates from the interference of the incident and a surface scattered wave. It is suggested that ultrafast laser induced surface modification in semiconductors is feasible in submelting conditions, and it may act as a precursor of the incubation effects observed at multiple pulse irradiation of materials surfaces.

# I.   INTRODUCTION

Silicon surface processing with ultra-short pulsed lasers has received considerable attention over the past decades due to its important applications in modern technology and medicine [1-6] . These abundant applications require a thorough knowledge of the laser interaction with the target material for enhanced controllability of the resulting modification of the target relief. Ultrashort laser heating process, in general, is pertinent to material removal and, in particular, ablation requires material irradiation with high intensity that induces mass removal [7-15]. Nevertheless, surface modification does not always entail mass removal through phase change (i.e. melting or evaporation). In particular, ultrafast laser processing with repetitive pulses of low fluence gives rise to incubation or accumulation effects [16]. This phenomenon is a direct consequence of the fact that following the first irradiating pulse a material already has changed into an often undefined condition. Due to such effects, for example, the coupling efficiency increases significantly with successive pulses in the case of multiphoton absorption [17]. It is thus anticipated that even sub-threshold multiple-pulse treatments have the potential to irreversibly modify materials chemically and mechanically. Therefore, the investigation of incubation effects is of overriding importance at least as a bridge between fundamental studies and applied materials science. An interesting effect which is the result of repetitive exposure to pulsed lasers is the ripple formation. Previous theoretical approaches or experimental observations related to these periodic structures were performed in melting [18] or ablation conditions [8, 10, 12, 19-21]. An alternative mechanism of ripple formation due to instabilities generated by the ultra-short pulses, similar to ion-beam sputtering has also been reported [22, 23],where laser induced instabilities resemble typical hydrodynamic instabilities of thin liquid films and laser-induced surface patterning is described by a non-linear surface erosion model. Nevertheless, the mechanism does not correlate the observed surface patterning with temperature gradients due to spatial dependent heat distribution and there is no prediction for pulse number relation to ripple periodicity values. In regard to the investigation of surface patterning in submelting conditions, to the best of our knowledge, there is no experimental evidence of ripple formation. Furthermore, a theoretical model that describes the mechanical response of the system due to ultrashort pulsed laser irradiation which also incorporates the surface signature ingredient is still elusive.

The present effort aims at the elucidation of the physical fundamentals and investigation of laser induced mechanical changes that can be attributed to incubation effects occurring at the submelting fluence regime. A rapid change in the lattice temperature can result in the development of thermal strains and volume expansion accompanied by the development of thermal stresses in the heated region. These, initially elastic stresses, may evolve to attain large values exceeding the yield stress of irradiated material, thus inducing dislocation motion that can affect performance in microelectronics [24, 25]. The purpose of elaboration of plastic deformation in submelting conditions is twofold: (i) to determine the conditions that govern elasticity, dislocation formation or plasticity and device failure, (ii) to elucidate previously unexplored aspects of material laser irradiation where it was assumed that surface modification occurs only in areas of the material where intensity of the laser beam is very high and sufficient to melt or evaporate the affected region.

Previous theoretical and experimental works on thermal stress development and structural modification are limited to metals irradiated by ultrashort-pulsed lasers [26-31], however, these studies focused mainly on plastic deformation induced by phase change where the complications of melting and/or evaporation limit the model predictions [32]. In particular, the response of silicon surfaces exposed to long pulsed lasers was considered and semi-analytical expressions for stress and strains were obtained [33, 34], by neglecting, however, dynamic plasticity effects. Moreover, the temperature dependence of the relevant thermophysical parameters (conductivity, heat capacity, absorption coefficient, etc) was ignored and its effects on plasticity were not taken into account. In this comparison, it is noted that although plastic deformation of silicon surfaces after irradiation with ultrashort pulses has been observed experimentally [35], no prior theoretical investigation has been conducted to probe the onset of the von Mises stress which defines the emergence of plastic deformation. Hence, the determination of the von Mises stress from an initial elastic distribution of stresses and strains within a dynamic thermoplasticity framework will allow a systematic analysis of the emergence of plasticity. Previous approaches required the incorporation of a plastic strain component  that is related to the velocity and density of dislocations [36]. The proposed model offers an alternative approach to describe plastic deformation which is free of the complexity presented by the need to estimate the above microscopic quantities, however, it predicts surface morphology tendency.

The purpose of this article is to outline a unifying theoretical photo-thermo-mechanical framework to describe the non-equilibrium energy transport between excited carriers and lattice coupled to the dynamic equations of equilibrium along with corresponding constitutive equations defining thermoelastic deformation, as well as, the onset of plasticity. In particular, a hybrid theoretical model is presented to illustrate a possible mechanism that leads to the emergence of plastic deformation upon laser irradiation in submelting conditions and crater formation. In addition, the evolution of plastic deformation in submelting conditions is considered by investigating a possible ripple formation after subsequent irradiation with sub-picosecond laser pulses. The model comprises of (i) a heat transfer component that accounts for a two-temperature heat conduction process coupled with elastic deformation, which also includes a thermal-mechanical coupling term, and, (ii) a dynamic thermoplastic component describing the evolution of the displacement field and associated thermal stress distribution, from which the von-Mises stress responsible for the onset of yielding can be computed. Details pertaining to the effect on the thermo-physical properties and the surface modification under the change of the laser beam fluence can be obtained. Due to the picometre size of the material morphology change for irradiation with one or a few pulses, an experimental validation of the theoretical model is not attainable and therefore a procedure to test the theoretical framework with experimental observations at measurable scales (i.e. nanosized structures) should be pursued by using a large number of pulses.

## II. MODEL

### 1. Irradiation with one pulse (NP=1)

Ultrashort-pulses lasers excite the energy carriers (electron-hole) in semiconductors and the energy is transferred to the lattice system. The relaxation time approximation to Boltzmann's transport equation [13] is employed to determine the number density, carrier and lattice energies. The evolution of the number density $N$, carrier temperature $T_c$ and lattice temperature $T_l$ are derived using the carrier balance, carrier energy balance and lattice heat balance. Based on this picture the following set of equations determine the temperature and particle dynamics

$$C_c \frac{\partial T_c}{\partial t} = \vec{\nabla} \bullet \left( (k_e + k_h) \vec{\nabla} T_c \right) - \frac{C_c}{\tau_e} \left( T_c - T_l \right) + S(\vec{r}, t)$$

$$C_l \frac{\partial T_l}{\partial t} = \vec{\nabla} \bullet \left( K_l \vec{\nabla} T_l \right) + \frac{C_c}{\tau_e} \left( T_c - T_l \right)$$

$$\frac{\partial N}{\partial t} = \frac{\alpha}{h\nu} \Omega I(\vec{r}, t) + \frac{\beta}{2h\nu} \Omega^2 I^2(\vec{r}, t) - \gamma N^3 + \theta N - \vec{\nabla} \bullet \vec{J} \qquad (1)$$

$$\Omega = 1 - R(T_l)$$

where $C_c$ and $C_l$ are the heat capacity of electron-hole pairs and lattice, respectively, $\nu$ is the frequency of the laser beam corresponding to a wavelength equal to $\lambda$=800nm, $k_e$ and $k_h$ are the thermal conductivity of the electrons and holes, respectively, $K_l$ is the thermal conductivity of the lattice, $h$ is the Plank's constant, $\gamma$ is the Auger recombination coefficient, $\theta$ is the impact ionisation coefficient, $\alpha$ and $\beta$ are the one-photon and two-photon absorption coefficients, $R(T_l)$ is the reflectivity of the laser beam on the silicon surface (which is a function of the material temperature), $\tau_e$ is the energy relaxation time, $\vec{J}$ is the carrier current vector and $S(\vec{r}, t)$ is provided by the following expression

$$S(\vec{r}, t) = (\alpha + \Theta N) \Omega I(\vec{r}, t) + \beta \Omega^2 I^2(\vec{r}, t) - \frac{\partial N}{\partial t} \left( E_g + 3k_B T_c \right) - N \frac{\partial E_g}{\partial T_l} \frac{\partial T_l}{\partial t} - \vec{\nabla} \bullet \left( \left( E_g + 4k_B T_c \right) \vec{J} \right) \qquad (2)$$

where $\Theta$ stands for the free-carrier absorption cross section, $k_B$ is the Boltzmann's constant and $E_g$ the band-gap energy. For the sake of simplicity, it is assumed that the incident beam is normal to a flat irradiated surface. The contribution of the current vector in the balance equation for the electron-hole carriers depends largely on the pulse duration. The laser intensity in Eqs .(1) and (2) is obtained by considering the propagation loss due to one- and two- photon absorption and free carrier absorption [13]

$$\frac{\partial I(\vec{r},t)}{\partial z} = -(\alpha + \Theta N)I(\vec{r},t) - \beta I^2(\vec{r},t) \tag{3}$$

On assuming that the laser beam is Gaussian both temporally and spatially, the laser intensity at the incident surface is expressed as form

$$I(r,z=0,t) = \frac{2\sqrt{\ln 2}}{\sqrt{\pi}\tau_P} E_p e^{-\left(\frac{r^2}{R_0^2}\right)} e^{-4\ln 2\left(\frac{t-t_0}{\tau_P}\right)^2} \tag{4}$$

where $E_p$ is the fluence of the laser beam and $\tau_p$ is the pulse duration (i.e. full width at half maximum), $R_0$ is the irradiation spot-radius (distance from the centre at which the intensity drops to $1/e^2$ of the maximum intensity); $t_0$ is chosen to be equal to $3\tau_p$. The choice of the value of $t_0$ is based on the requirement that at $t=0$, the intensity of the incident beam is practically zero, at $t=t_0$ it reaches the maximum power while laser irradiation vanishes at $t=2t_0$.

The mechanical response of the material is described by the differential equations of dynamic equilibrium. For the present axially symmetric configuration, the dynamic equilibrium equations in cylindrical coordinates [37]

$$\rho \frac{\partial^2 u_r}{\partial t^2} = \frac{\partial \sigma_{rr}}{\partial r} + \frac{\partial \sigma_{rz}}{\partial z} + \frac{\sigma_{rr} - \sigma_{\theta\theta}}{r}$$
$$\rho \frac{\partial^2 u_z}{\partial t^2} = \frac{\partial \sigma_{zz}}{\partial r} + \frac{\partial \sigma_{zz}}{\partial z} + \frac{\sigma_{rz}}{r} \tag{5}$$

where $u_r$ and $u_z$ are the displacements in the $r$- and $z$-directions, respectively. The corresponding stresses $\sigma_{ij}$ ($i,j=r,\theta,z$) and strains $\varepsilon_{ij}$ in radial coordinates are assumed to be given initially by the equations of linear thermoelasticity [37]

$$\sigma_{rr} = \lambda\left(\varepsilon_{rr} + \varepsilon_{\theta\theta} + \varepsilon_{zz}\right) + 2\mu\varepsilon_{rr} - \frac{E}{1-2\nu}\alpha'(T_l - T_0)$$

$$\sigma_{\theta\theta} = \lambda\left(\varepsilon_{rr} + \varepsilon_{\theta\theta} + \varepsilon_{zz}\right) + 2\mu\varepsilon_{\theta\theta} - \frac{E}{1-2\nu}\alpha'(T_l - T_0)$$

$$\sigma_{zz} = \lambda\left(\varepsilon_{rr} + \varepsilon_{\theta\theta} + \varepsilon_{zz}\right) + 2\mu\varepsilon_{zz} - \frac{E}{1-2\nu}\alpha'(T_l - T_0) \tag{6}$$

$$\sigma_{rz} = \mu\varepsilon_{rz}$$

$$\varepsilon_{rr} = \frac{\partial u_r}{\partial r}, \varepsilon_{\theta\theta} = \frac{u_r}{r}, \varepsilon_{zz} = \frac{\partial u_z}{\partial z}, \varepsilon_{rz} = \frac{\partial u_r}{\partial z} + \frac{\partial u_z}{\partial r}$$

$$\lambda = \frac{E\nu}{(1+\nu)(1-2\nu)}, \mu = \frac{E}{2(1+\nu)}$$

where $E$ and $\nu$ are the Young modulus and Poisson ratio (for silicon, $E$=150GPa and $\nu$=0.17 [38]), respectively. Furthermore, $\rho$ is the density of the silicon (=2330Kgr/m³), ($T_l - T_0$) is the relative lattice temperature with respect to the ambient temperature ($T_0$ =300K) and $\alpha'$ is the coefficient of linear thermal expansion which , for temperatures below the melting point, is provided by the expression [39]

$$\alpha'(T_l) = \left[3.725\left(1 - e^{-5.88\times10^{-3}(T_l - 124)}\right) + 5.548\times10^{-4}T_l\right]\times10^{-6} \quad (\text{K}^{-1}) \tag{7}$$

which, in fact, is the departure from standard linear thermoelasticity where $\alpha'$ is assumed as a constant (for small temperature variations). In regard to the obvious additional complexity of the mechanical response of the material due to fact that silicon is an anisotropic medium, the computed picosecond size of plastic deformation allows an isotropic treatment of the thermomechanical response of the system at a first approximation [34]. An axisymmetric Gaussian heat source and the assumption of an approximately isotropic material allow a simplified 2D modelling of the thermomechanical response when a flat surface is irradiated.

The corresponding heat conduction equation reads [30]

$$C_l \frac{\partial T_l}{\partial t} = \vec{\nabla} \bullet \left( K_l \vec{\nabla} T_l \right) + \frac{C_e}{\tau_e} \left( T_e - T_l \right) - \frac{E}{1-2\nu} \alpha' T_0 \left( \varepsilon_{rr} + \varepsilon_{\theta\theta} + \varepsilon_{zz} \right) \qquad (8)$$

which modifies the second equation in Eqs.1 to include the effect of strain and also modifies the corresponding equation of linear thermoelasticity in a robust manner by allowing the aforementioned nonlinear temperature dependence of the thermal expansion coefficient $\alpha'$. To determine the onset of plastic deformation, the von Mises or effective stress ($\bar{\sigma}$)

$$\bar{\sigma} = \frac{1}{\sqrt{2}} \sqrt{\left( \sigma_{rr} - \sigma_{\theta\theta} \right)^2 + \left( \sigma_{\theta\theta} - \sigma_{zz} \right)^2 + \left( \sigma_{rr} - \sigma_{zz} \right)^2 + 6\sigma_{rz}^2} \qquad (9)$$

should be computed. When the value of $\bar{\sigma}$ reaches the value of the yield stress $\sigma_y$, plastic flow occurs. Due to the high strain rates involved, a departure from classical plasticity is adopted by assuming that the yield stress of the material depends on the strain rate ($\dot{\varepsilon}$) and temperature as follows [33, 34]

$$\sigma_Y(T_l) = C\left( \dot{\varepsilon} \right)^{1/n} \exp(\frac{U}{2k_B T_l}) \qquad (10)$$

where $U$ is the activation energy (for dislocation glide), $n$ and $C$ are constants depending on the material (for silicon, $n$=2.1, $U$=2.3eV, $C$=4.5x10$^5$ dyn/cm$^2$) [34].

## 2. Irradiation with more pulses ($NP$>1)

Irradiation of silicon with sufficient energy to induce surface deformation will eventually lead to a crater formation with a radially dependent depth upon heating with a single pulse. Further laser irradiation of the nonflat surface may give rise to a surface scattered wave [10] (Fig.1). According to existing theoretical predictions and experimental studies, the interference of the incident and the surface waves results in the development of subwavelength 'ripples' with orientation perpendicular to the electric field of the laser beam [10, 12, 21]. The ripple periodicity is given by $\lambda/(1+\sin\theta)$ [10, 21] where $\theta$ is the angle of the incident beam with respect to the normal axis on the irradiated surface. Due to the crater, subsequent pulses will not be normal to the new surface profile. An alternative scenario based on the interference of the incident wave with surface plasmons has been proposed [20] to explain the ripple formation on metal and semiconductors. A necessary condition for the development of surface plasmons on the semiconductor is a semiconductor-to-metal transition that occurs only if laser irradiation melts the material. The assumption that the irradiation conditions do not suffice to produce a solid-to-liquid phase transitions suggests that ripple formation could be attributed to the interference of the incident wave with a surface scattered wave (Fig.1). The average intensity of the superposition of the two waves on the surface of the material is derived by computing the Poynting vector and integrating it over the laser period ($\lambda/c$). Then, the final intensity on the surface is provided by the following expression

$$I_{surf}(\vec{r},t) = \left\langle \left| \vec{E}_i + n_{mat}\vec{E}_s \right|^2 \right\rangle e^{-\left( \frac{r^2}{R_0^2} \right)} e^{-4\ln 2 \left( \frac{t-t_0}{\tau_p} \right)^2} \qquad (11)$$

where $n_{mat}$ is the refractive index of silicon, $\vec{E}_i = \vec{E}_{i,o}(\vec{r}) \exp(-i\omega_i t + i\vec{k}_i \bullet \vec{r})$ and $\vec{E}_s = \vec{E}_{s,o}(\vec{r}) \exp(-i\omega_s t + i\vec{k}_s \bullet \vec{r})$ and $\omega_i$, $\omega_s$ are the frequencies of the incident beam (equal to $\omega$) and the surface scattered wave, respectively. The magnitude of the electric field of the incident wave can be calculated by the expression

$$\frac{E_d}{\tau_p} = \frac{c\varepsilon_0 \sqrt{\pi} \left\langle \left| \vec{E}_i \right|^2 \right\rangle}{2\sqrt{\ln 2}} \qquad (12)$$

while the magnitude of the electric field of the surface scattered wave (Fig.1) is taken to be of the order of the electric field of the incident wave. The computation of the time averaged quantity in the final intensity on the surface (Eq.11), yields a contribution proportional to $\cos\left(\sqrt{\vec{k}_i{}^2 + \vec{k}_s{}^2 - 2\vec{k}_i \bullet \vec{k}_s}\left(\cos\varphi\right)x\right)\vec{E}_{i,o} \bullet \vec{E}_{s,o}$ (x-axis is along the direction of the electric field of the incident beam, see Simulation Section), where $\vec{k}_i$ and $\vec{k}_s$ are the wavevectors of the incident and surface scattered waves, respectively. The above expression indicates that the polarisation of the incident beam is crucial for the determination of the form of the total intensity distribution. Due to the curvature of the initially modified surface after irradiation with one laser pulse, the direction of the surface scattered waves is geographically correlated. Nevertheless, the energy deposition will be highest in the direction of the laser electrical field and lowest in the perpendicular direction because the electric field of the surface scattered wave has a larger component along the polarisation of the incident beam. Hence, the energy deposition is strongly correlated to the polarisation of the laser beam and it yields a highest component along the incident electric field polarisation. As a result, for an incident beam with polarisation on the xz-plane (Fig.10a), the periodic function produces an optical interference pattern which propagates in parallel to the polarisation vector and it is followed by a spatially and periodically modulated energy deposition. In the next sections, it will be demonstrated that a periodic distribution of energy will lead to an alteration between compressive and tensile axial regimes which yields depression and rise of surface, respectively, which is a similar scenario to capillarity-driven effects proposed in melting conditions [18]. As a time dependent deformation of the surface profile is expected to occur due to exposure to each pulse, it is important to emphasise that subsequent pulses are assumed to irradiate material which practically ceased to undergo a deformation. This is a correct assumption if the time gap between subsequent pulses is larger than 1 msec and then, Eq. (11) is still valid.

### III. SIMULATION

Due to the temperature dependence of the thermo-physical parameters and the strong nonlinearity of the thermomechanical equations, an analytical solution of the coupled system of equations (Eqs.1-8) cannot be pursued. Furthermore, conditions that lead to thermally induced displacements entail a deformed mesh and thereby a solution by means of a finite difference method is expected to produce questionable results and therefore a finite element method is employed. Due to the radial dependence of the Gausian beam, an axisymmetric modelling is performed to investigate the thermal response of ultrashort laser pulse interaction with the silicon film and the thermomechanical response and displacement (for single pulse irradiation). By contrast, interference of the surface scattered waves with the incident beam breaks the initial axial symmetry of the system and thereby a three dimensional procedure should be ensued to analyse the process for irradiation with subsequent pulses. In that case, the coordinate system used in the analysis is defined as follows: the z-axis is normal to the material surface, the x-axis is on the surface with a direction based on that the electric field of the incident beam must reside on the xz-plane, and the y-axis, again on the material surface (Fig.10a).

Simulations were conducted for various values of the fluence $E_p$ (in the range 0.05-0.25J/cm$^2$, see explanation of the choice in the *Results & Discussion* section) and pulse duration ($\tau_p$=430fs) to examine the dependence of the effects on laser characteristics. Eqs. (1-8) are solved along the radial and axial directions to obtain the carrier and lattice temperatures, as well as, the surface modification details. The temperature-dependent thermophysical parameters that are used in the numerical solution of the governing equations are listed in Table I. The top boundary was discretised with a maximum element size set equal to 10pm while the maximum element size for the rest of the subdomain was set to 50pm. Moreover, the region where deformation effects may not be important enough has been discretised with a maximum element size equal to 20nm. A time-dependent solver was used together with a direct linear system solver. Carriers and lattice temperatures are set equal to $T=T_0$=300$^0$K (room temperature) at $t$=0 while the initial concentration of the carrier is set to $N$=1μm$^{-3}$ [13]. During the ultrashort period of laser heating, heat loss from the upper surface of target is assumed to be negligible. As a result, the heat flux boundary conditions for carriers and lattice are zero throughout the simulation. Furthermore, it is assumed that only the top surface is subjected to the Gaussian-shape laser beam of an irradiation spot-radius $R_0$=7.5μm. To simulate the mechanical response, Eq.5 is used, whereas only vertical displacements are allowed along the symmetry axis ($r$=0); negligible total displacement and stresses are assumed far from the area affected by the laser beam. The same is pursued for the azimuthal component of the stress along the symmetry axis $\sigma_{\theta\theta}$(r=0, z, t) and the axial stress components on the surface and the bottom side, $\sigma_{zz}$(r, z=0, t) and $\sigma_{zz}$(r, $z_{bottom}$, t), respectively, are set to zero. For $NP>1$, a special attention is required to take into account the modified surface profile. The absorbed energy is directly dependent on the angle at which the incident beam irradiated the material. Although an approximate flat surface may be assumed for low number of pulses, a correction due to a radially dependent depth is necessary to attain accurate results as irradiation shots increase. Hence, typical Fresnel equations are used to describe the reflection and transmission of the incident light while $R_{new}$ and $Z_{new}$ represent the tangential and vertical axes on a moving reference frame, respectively.

## IV. RESULTS AND DISCUSSION

The results presented below were produced by irradiation with a laser beam of peak fluence $E_p$=0.13J/cm$^2$ which is not sufficient to melt the material, however, the applied intensity can produce thermally induced stresses that lead to plastic deformation. According to the simulation results, the maximum lattice temperature under the aforementioned conditions after equilibration is $T$=950$^0$K, significantly below the silicon melting point. A simultaneous solution of the heat transfer and dynamic thermoelasticity equations yields the spatio-temporal distribution of stress, strains and elastic displacement of the lattice nodes. Fig.2a illustrates the spatial distribution of the $r$-component of the stress, $\sigma_{rr}$, and, in particular, the radial profile is illustrated at three locations in the bulk at $t$=1ns. According to Fig.2b, the stress is compressive close to the surface ($Z$=-0.5µm) and this trend for the stress continues with a smaller magnitude as the depth increases ($Z$=-1.5µm). This magnitude decrease is expected as the temperature field decays in the $z$-direction and it tends gradually to zero far away from the surface into the bulk. Furthermore, Fig.2b indicates that at large depths, the stress changes from compressive to tensile with a substantially smaller magnitude ($Z$=-6.5µm). All profiles show that the gradient of lattice temperature in the $r$-direction governs the spatial decrease noted in these curves. Similar results were attained for the radial stress component after irradiation of metal thin films with short pulsed lasers [31].

The spatial distribution of the axial stress, $\sigma_{zz}$ is illustrated in Fig.3a while Fig.3b shows the $z$-dependence of the stress profiles at three different locations ($r$=2, 6, 10µm) at $t$=1ns. According to Fig.3b, a similar trend to that of the radial stress is followed by the axial stress. More specifically, axial stress is compressive at distances close to the symmetry axis ($r$=0) with a decreasing magnitude at increasing radial distances. Interestingly, the axial stress component is tensile at distances larger than the irradiation spot radius $R_0$ ($r$=10µm). Moreover, similarly to the results for the radial stress, there is an indication for significant differences between the maximum magnitudes of the compressive and tensile stresses.

Although the above exploration of the distribution of the stresses and associated strains provides information on the initial elastic response of the material, determination of emergence of plasticity and irreversible deformation results from considering the spatio-temporal evolution of the von Mises stress (Eq.9). Fig.4a illustrates the distribution of the effective stress at $t$=1ns and profiles at different axial and radial positions have been computed. The profile of the effective stress reveals the significance of the component stresses in the decrease of the von Mises stress at increasing distance from the centre of the irradiated spot. The decrease of the effective stress along the radial direction follows the profile of the temperature field in the same direction (Fig.4b). Similarly, a more abrupt temperature decrease in the axial direction inflicts a similar behaviour on the von Mises stress (Fig.4c).

Next it is noted that existing experimental data can allow the estimation of the temperature dependence of the (lower) yield stress through a logarithmic expression [40]. In the present study, this formula was used to evaluate yield stress at all lattice temperatures and compute its spatio-temporal distribution of the yield stress. Fig.4d illustrates the spatial distribution of the yield stress at $t$=1ns which exhibits a magnitude increase with increasing distance from the centre of the irradiated spot, as expected due to the lower lattice temperature gradients involved. The spatio-temporal distribution of the axial displacement has been computed (Fig.5a) and indicating that its magnitude is pronounced not only in accordance to the Gaussian profile of the laser but also decreasing in accordance with the fluence ($\sim$exp(-r$^2$/(R$_0$)$^2$)), see Fig.5b. Furthermore, Figs.5a,b indicate that the increase in the compressive radial stress with decreasing vertical distances from the irradiated region centre leads to a surface suppression and a crater formation (without, however, rigorously accounting separately for elastic, as well as, plastic effects). To ascertain the conditions of the emergence of plasticity, the spatio-temporal distribution of the von Mises stress is required. Fig.6a illustrates the spatial stress distribution of the von Mises stress at $t$=1ns. The region inside the boundary (*white* line) corresponds to values higher than the yield stress which is necessary for plastic deformation. It appears that only a small area of size approximately half that of the laser beam spot (i.e. ~3.4μm) is plastically deformed while the rest is characterised by elasticity. Nevertheless, despite the elastic behaviour in regions outside the boundary, the surface profile will indirectly be modified outside the plastically affected region due to the constrained motion inflicted by the plastic deformation experienced by the former area. As a result, an extended area will undergo a plastic deformation despite the associated stress does not exceed the yield stress at distances larger than 3.4μm. To avoid confusion, it is important to note that the profile suppression is related exclusively to movement of the lattice points and an elasticity-to-plasticity transition and no mass removal governs the surface modification. Figure 6b illustrates such a profile modification at $z$=0.1μm at $t$=1ns and $t$=100ns which confirms the above mentioned yielding at distances less than 3.4μm and indirect plasticity due to constrained motion at larger distances. The choice of the timepoint, $t$=100ns, was based on the fact that at later stages, mechanical response is minimal and associated plastic effects can be ignored. The minimal mechanical response at later stages justifies the earlier assumption that further irradiation will heat an already deformed material provided that the temporal gap between the two pulses is larger than 1ms.

To determine the influence of the laser beam characteristics on the size of the crater depth, simulations are performed for various fluences ($E_p$=0.05J/cm$^2$ to $E_p$=0.25J/cm$^2$) and the results are illustrated in Fig.7a. It appears that the vertical plastic displacement follows the fluence monotonically due to the fact that larger values result to higher temperatures and stresses. Further investigation of the maximum displacement dependence on the characteristics of the laser beam demonstrate that fluence values lower than 0.05J/cm$^2$ induce very small displacements which, in regard to the simulation, are within the computational error. By contrast, increase of the fluence leads to larger displacement values of the deformation (Fig.7a). The ability to determine the correlation between the size of deformation and the fluence indicates that regulation of the deformation in the material and plasticity prediction can be achieved by modulating the laser beam characteristics. Similar prediction is anticipated through modulation of the rest of the laser beam characteristics (i.e. pulse duration, radial spot size). Special attention is required for fluences that exceed the value $E_{pm}$=0.25J/cm$^2$ and present limitations to the proposed physical mechanism. Firstly, excessive load generated by high fluences can induce fracture which means an alternative mechanism is required to be introduced. Secondly, at high fluences, material melts and the presence of molten phase limits the validity of the presented theoretical framework to describe the underlying physical mechanism of deformation. The behaviour can be explained by the fact that the phase transformation alters the properties of the material and a rigorous theoretical approach should incorporate hydrodynamics (for molten phase) and yield stress changes after resolidification. Although previous works that aimed to describe microbump formation in thin metal films were based on the analysis of similar thermomechanical equations [27], a revised and more complex version of the presented model is required to incorporate the above processes for plastic deformation for fluences larger than $E_{pm}$. Furthermore, the employment of the model for conditions that leads to melting also suggests volume suppression and precise and convincing conclusions are debatable. This is due to the fact that a generalised theory which combines elasticity and flow effects is required to simulate and characterise the shear flow of materials in a molten phase. As a result, a complete method that takes into account both stress related displacement and alteration of downward and upward movement is difficult to be introduced for a material that undergoes phase transition. In this connection, it is noted that a combination of a two temperature model and molecular dynamics [41], may potentially present an alternative technique to compute deformation characteristics and related spatio-temporal pattern forming instabilities. Similarly, gradient plasticity theories [42-45] may provide an improved methodology to address such deformation instability phenomena by combining a higher-order two-temperature model [42] (describing the nonequilibrium carrier-lattice thermal relaxation process) with a higher-order gradient plasticity model (describing the heterogeneous character of plastic flow).

In relation to the above discussion it is noted that the proposed model aims at predicting ripple formation (when the material is further irradiated) due to the interference of the incident with surface scattered waves within a much less sophisticated mechanical framework. In fact, it has already been mentioned that subsequent irradiation increases the maximum surface depth, while a similar behavior (monotonicity) is ensued by the ripple height. The maximum number of pulses performed in the simulation for $E_p$=0.13J/cm$^2$ was $NP$=1100; for larger number of pulses the effective stress values exceed the threshold for emergence of fracture (i.e. fracture occurs when stress exceeds 7GPa [46]).

Fig.7b illustrates the spatial dependence of the ripple profile on the *xz*-plane for 600, 800, and 1000 pulses at *t*=100ns. Interestingly, an increase in the number of pulses results into a larger spot radius which has been observed under ablation conditions [8, 11]. With respect to surface modification induced in submelting conditions, the monotonicity is justified by the increase of the radial extent of the vertically displaced material with increasing *NP* due to the constrained motion. Simulation results indicate that the maximum size of deformation that corresponds to maximum peak-to-valley height is 1.56nm for 1000 pulses. This indicates that although a single pulse is capable to produce subpicometre-size morphology change, repetitive irradiation is capable to further modify surface profile and induce nanosize structures in submelting conditions. Furthermore, ripple periodicity for *NP*=1000 is characterised by an average value equal to 718nm which suggests that repetitive irradiation of the material produces subwavelength ripples. According to the fluence intensity and corresponding dependence surface modification results (Fig.7a), larger nanostructures can be produced, the size of which can be controlled by appropriate variation of the laser beam power. Fig.8α illustrates the theoretically predicted three dimensional spatial dependence of the surface patterning for *NP*=1000 in one quadrant where there is a pronounced rippled surface. It is evident that although the rippled surface protrusions are above the initially unaffected surface at *z*=0 (Fig.7b), at *y*=1.84μm some tips of the plastic deformed surface are below the flat surface (Fig.8b). This behaviour is due to the fact that at larger distances, energy deposition is smaller which leads firstly to smaller temperature gradients and subsequently to stress and strains differences.

Given the maximum depth and periodicity of the ripples dependence on the number of pulses (Fig. 7b) we performed a thorough investigation on the correlation of these parameters in the whole range of pulse numbers. It is evident that the maximum depth of the plastically deformed region increases when the material surface is exposed to more pulses (Fig.9a). Furthermore, we notice that for large *NP*, ripples develop above the initial level of the surface profile (i.e flat surface) while for smaller values of *NP*, ripples form below *z*=0 (Fig.7b). This is due to the fact that for small *NP* stress gradients are not large enough to produce significant alteration between tensile and compressive regimes. As a result, the associated surface depression and protrusions will not be very pronounced for small *NP*. By contrast, a decrease of the ripple periodicity is observed with increasing number of pulses (Fig.9b); this behaviour has also been exhibited by semiconductors, metals and dielectrics in ablation or subablation conditions [10, 19, 20, 47].

To validate experimentally the theoretical results, experiments involving irradiation of a single crystal (100) Si wafer surface with a femtosecond laser beam were performed. The aim was to obtain some initial experimental support on the soundness of the model assumptions used, rather than a detailed quantitative checking on model predictions. The latter may be a subject for the future when gradient plasticity and instability deformation analysis could be incorporated in the theoretical study in an efficient manner. Furthermore, a thorough investigation would entail on the consideration of the anisotropic mechanical behaviour of silicon. Finally photochemical decomposition leading to bond-breaking may be considered to provide an additional contribution, however, its role is not expected to be very significant [48]. Nevertheless the aim of the current work aimed at providing a simple picture of the surface signature that resembles the material profile in a first approximation. To test the theoretical framework, the laser used was a Ti:sapphire regenerative amplifier source with wavelength and repetition rate of 800 nm and 1 kHz, respectively. The pulse duration was set to 430 fs and measured by means of cross correlation techniques. The laser beam was focused onto the sample, located inside a vacuum chamber evacuated down to a residual pressure of $10^{-2}$ mbar, by means of a plano-convex lens of 100 mm focal length. The number of irradiating pulses was controlled by a mechanical shutter. All experiments were performed at the fluence of 0.13J/cm², which is well below the modification threshold of the material. Following laser irradiation, the sample surface was observed by a field emission scanning electron microscope (FESEM, JEOL 7000). Fig.10a illustrates a top-view SEM image of the spot attained after irradiation with 1000 pulses and the intensity profile along a perpendicular to ripples direction (*dashed* line) was used to estimate ripple periodicity. It should be noted that this profile provides accurate information only for the dimensions on the sample plane, while in the vertical direction it gives relative rather than the absolute dimensional values. It is evident that for a *p*-polarised beam the ripples develop perpendicularly to the electric field of the incident beam which also has been reported in laser induced welding, hardening and annealing [8, 10, 21]. Image analysis techniques including a fast Fourier transform (FFT) low pass filter were performed to remove noise and produce the intensity profile without significant fluctuation. Estimation of ripple periodicity is achievable through the computation of the horizontal distance between the pronounced local minima (*circles* in Fig.10b) which yields an average experimental value of 700±40nm. A more systematic approach through a frequency analysis of the SEM image and plot of the complex magnitude of the FFT of the image intensity squared against the period yields a pronounced peak at 730nm (Fig.10c). The observed discrepancy between the experimental ripple profile and the theoretical predictions for $NP$=1000 can be attributed partly to a possible experimental inaccuracy (i.e. fluence, spatio-temporal resolution of the laser beam, material defects), but also to the simplifications introduced into the theoretical model employed. Nevertheless, the resulting ripple profile and average measured periodicity is in within the predictions of the theoretical model for $NP$=1000 (i.e. 718nm) which suggests that the assumptions incorporated in the proposed theoretical framework are in the right direction for predicting the formation of ripples and account satisfactorily for incubation effects. A significant aspect which needs to be emphasised is that the ripple periodicity is controlled by the mechanical response of the system. The morphology and the surface profile of the material before irradiation by subsequent pulses is determined through a dynamic thermoelasticity stress-analysis, also used to determine the von Mises stress responsible for yielding. Therefore, the ripple periodicity is directly related to the laser induced thermoplastic surface deformation and the agreement between theory and experiment indicates that the model appears to describe adequately the mechanism that induces incubation effects. However, despite a satisfactory interpretation of effects that characterise the submelting regime, a more thorough experimental investigation is required to deduce more details about the response of the system in various submelting conditions (i.e. different fluence, different laser beam characteristics and different number of pulses), which in turn can be used for model refinements. Accordingly, an extensive analysis of the influence of various parameters in the alteration of ripple height and periodicity is beyond the scope of the present work. Furthermore, a test of the influence of a variety of conditions on the surface modification requires an accurate methodology for measuring experimentally the spatial characteristics and intensity of the induced maximum plastic deformation. Due to the small vertical plastic displacement that ranges from some picometres to an approximately 1.56nm, AFM techniques may not be capable to precisely measure the deformation as it is in within the experimental error of the apparatus. However, recently developed techniques that have been proposed to obtain accurate observations in experiments on this scale are based either on X-ray [49] or femtosecond electron diffraction [50] methods, may be promising tools to facilitate the experimental validation of such investigations.

The new feature of the proposed model is the inclusion of thermally induced plastic effects. It was shown that the increased size of the affected area due to heating under the laser irradiation conditions used, may not be interpreted via other conventional physical mechanisms, such as melting [51, 52] or ablation [26], which do not sufficiently quantify the measured dimensions of the resulting spot size. Although the proposed scenarios may adequately describe, at least qualitatively, the underlying physical processes, the assumption that the related phenomena are confined in a region inside the laser beam spot area underestimates the contribution of the thermal stress deformation. According to our analysis, temperature gradients are the key factor for the development of stresses that can induce plastic deformation. Given the Gaussian intensity profile, higher values of peak fluence that are capable to induce ablation or melting suffice to create large temperatures (below the melting point) and associated gradients even outside the laser beam spot area. Hence, the associated plastic modification details may not be ignored and the resulting vertical displacement needs to be considered. Thus, the inclusion of thermal stress induced deformation appears to be an important ingredient in any temperature-gradient related surface modification mechanism. In fact, it may be advanced that the presented model could provide a methodology for investigating the processes that characterise plastic deformation in submelting conditions, but also for establishing the underlying mechanisms of material nanopatterning (e.g. nanocracks, nanoholes) in the submelting threshold regime. This may potentially open the way for developing new technologies and novel tools for surface nanopatterning based on laser induced stress fields. Furthermore, thermally induced dislocation motion and damage evolution can potentially be systematically analysed, in principle within the proposed framework. Elucidation of the conditions that lead to plasticity and damage is an aspect in semiconducting materials. Elimination of plastic deformation and damage can enhance functionality and performance of semiconducting components in a wide range of applications (microelectronics, photonics, medical devices, etc.). Moreover, the proposed methodology can be generalised and applied to different classes of materials such as metals and dielectrics by considering appropriate alterations in the laser-matter heat transfer mechanism. More specifically, elucidation of the mechanical properties of nanowires after irradiation with ultrashort pulses is extremely important for emerging applications in the areas of nanomaterials, nanocomposites and nanoelectromechanical devices. Similarly, structural modification of fused silica after irradiation can induce refractive index changes and alteration of optoelectronic properties of optical components.

## V. CONCLUSIONS

We have presented a detailed theoretical hybrid model to describe the physical fundamentals of plastic deformation due to thermal stresses during semiconductor irradiation with ultrashort laser pulses in submelting conditions. It contains elements of laser beam characteristics, heat transfer processes, as well as thermoelasticity and plastic yielding considerations that describe the particle dynamics and conduction phenomena, as well as the induced deformation processes that lead to surface morphology modification. By choosing appropriate values for the beam fluence, we demonstrated that it is likely to modulate the surface of the material and the magnitude of the elated plastic deformation was deduced by comparing the induced von Mises stress with the yield stress of the material. Characterisation of the conditions that lead to crater development and ripple formation was addressed and some experimental evidence on the soundness of the proposed model was provided.

### Acknowledgements

The authors gratefully acknowledge financial support from KEA's ERC Starting Grant MINATRAN 211166. The authors would like also to thank E.C. Aifantis for a series of useful discussions.

| Quantity | Symbol (Units) | Value |
|---|---|---|
| Initial temperature | $T_0$ ($^0$K) | 300 |
| Electron-hole pair heat capacity | $C_c$ (J/$\mu$m$^3$ K) | $3Nk_B$ |
| Electron-hole pair conductivity [53] | $K_c$ (W/$\mu$m K) | $10^{-6} \times (-0.5552 + 7.1 \times 10^{-3} \times T_c)$ |
| Lattice heat capacity [54] | $C_l$ (J/$\mu$m$^3$ K) | $10^{-12} \times (1.978 + 3.54 \times 10^{-4} \times T_l - 3.68\, T_l^{-2})$ |
| Lattice heat conductivity [54] | $K_l$ (W/$\mu$m K) | $0.1585\, T_l^{-1.23}$ |
| Band gap energy [55] | $E_g$ (J) | $1.6 \times 10^{-19} \times (1.167 - 0.0258\, T_l\,/T_0 - 0.0198\, \left(T_l\,/T_0\right)^2)$ |
| Interband absorption (800nm) [55] | $\alpha$ ($\mu$m$^{-1}$) | $0.112\, e^{T_l/430}$ |
| Two-photon absorption (800nm) [55] | $\beta$ (sec $\mu$m/J) | $9 \times 10^{-5}$ |
| Reflectivity (800 nm) [56] | $R$ | $0.329 + 5 \times 10^{-5}\, T_l$ |
| Latent heat of melting [57] | $L_m$ (J /$\mu$m$^3$) | $4206 \times 10^{-12}$ |
| Melting temperature [57] | $T_m$ (K) | 1687 |
| Auger recombination coefficient [13] | $\gamma$ ($\mu$m$^6$/sec) | $3.8 \times 10^{-7}$ |
| Impact ionisation coefficient [13] | $\theta$ (sec$^{-1}$) | $3.6 \times 10^{10}\, e^{-1.5 E_g /k_B T_c}$ |
| Free carrier absorption cross section (800nm) [55] | $\Theta$ ($\mu$m$^2$) | $2.9 \times 10^{-10}\, T_l\,/T_0$ |
| Energy relaxation time [53] | $\tau_e$ (sec) | $\tau_{e0}\left[1 + \left(\dfrac{N}{N_{cr}}\right)^2\right]$, $\tau_{e0} = 0.5 \times 10^{-12}$ sec, $N_{cr} = 2 \times 10^9\, \mu$m$^{-3}$ |

**Table I:** **Model parameters for Si.**

Fig. 1: Irradiation of a non-flat Si-surface with a Gaussian beam. Incident (denoted by the index *i*) and surface scattered waves (denoted by the index *s*) are depicted in the inset. Wave vectors and direction of the electric field of the two waves are sketched.

Fig. 2 (Color online): Stress component $\sigma_{rr}$ at *t*=1ns (a). Radial dependence of $\sigma_{rr}$ at three different locations inside the material is also illustrated (b).

Fig. 3 (Color online): Stress component $\sigma_{zz}$ at *t*=1ns (a). Axial dependence of $\sigma_{zz}$ at three different values of the radial distance is also illustrated (b).

.

Fig. 4 (Color online): (a). Von Mises stress component $\overline{\sigma}$ at *t*=1ns. (b). Radial dependence of $\overline{\sigma}$ at three different locations inside the material. (c). Axial dependence of $\overline{\sigma}$ at three different locations inside the material. (d). Yield stress spatial distribution at *t*=1ns.

Fig. 5 (Color online): (a). Spatial distribution of vertical displacement at *t*=1ns. (b). Vertical displacement dependence on the axial distance.

Fig. 6 (Color online): (a). Illustration of Mises stress component $\overline{\sigma}$ at *t*=1ns. Region inside the white border is plastically deformed. (b). Spatial distribution of vertical displacement at *t*=1ns and *t*=100ns for *NP*=1.

Fig. 7 (Color online): (a). Correlation between maximum vertical displacement and fluence values in submelting conditions for *NP*=1 for a Gaussian beam of peak fluence $E_p$=0.13J/cm², (b). Spatial distribution of vertical displacement after irradiation with a Gaussian beam of peak fluence $E_p$=0.13J/cm² at *t*=100ns for *NP*=600, 800, 1000 (*xz*-plane profile).

Fig. 8 (Color online): (a) Ripple formation (simulation results) for *NP*=1000, (b) Spatial distribution of vertical displacement at *y*=1.84μm.

Fig. 9: (a). Correlation between maximum vertical displacement and number of pulses, (b). Dependence of ripple periodicity on the number of pulses.

Fig. 10: (a). SEM image after irradiation with 1000 pulses (*p*-polarised beam), (b). Intensity profile across dashed white line in the previous image (*circles* illustrate the local minima of the ripples). (c). Fast Fourier transform of the intensity image and computation of most pronounced period.

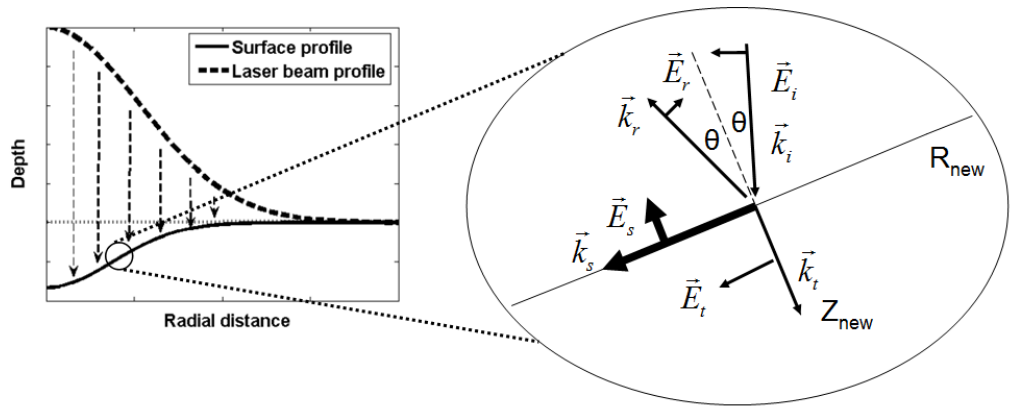

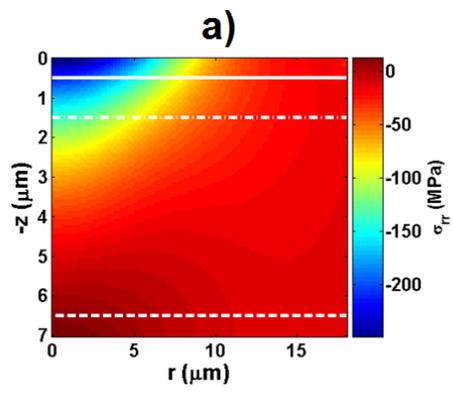
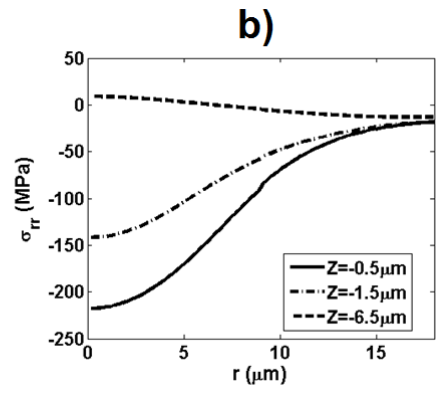

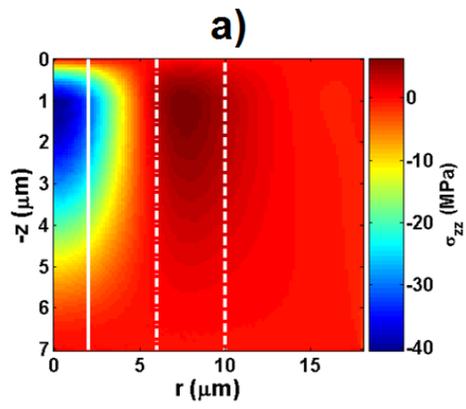

**a)**

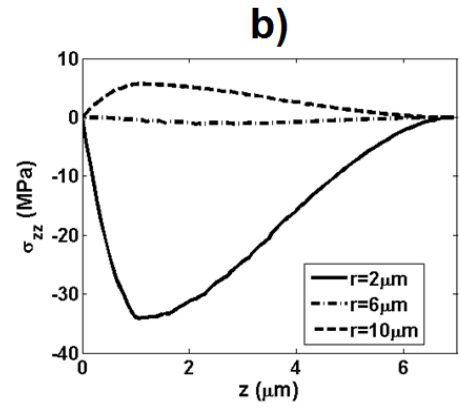

**b)**

**a)**

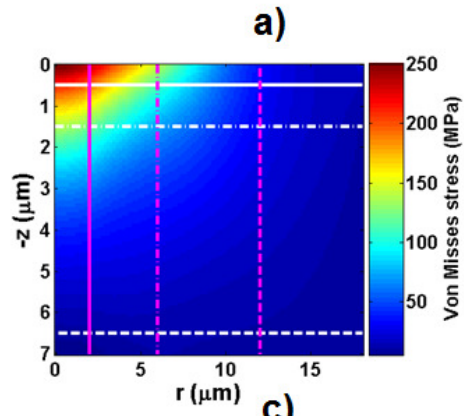

**b)**

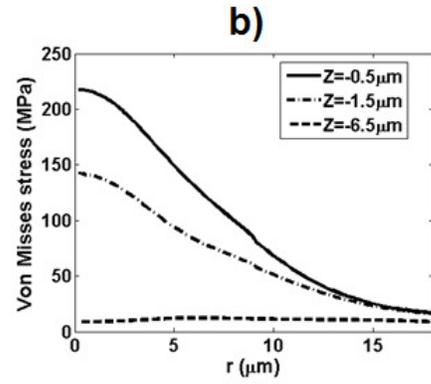

**c)**

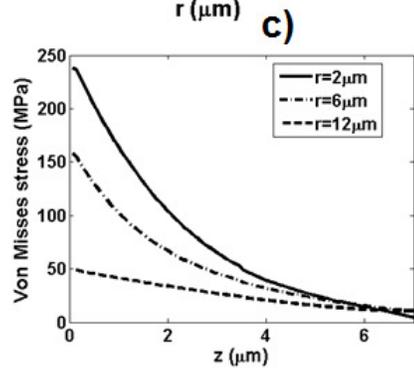

**d)**

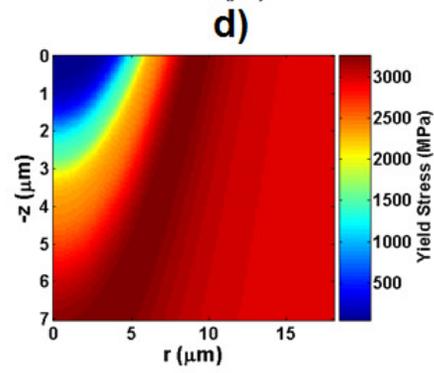

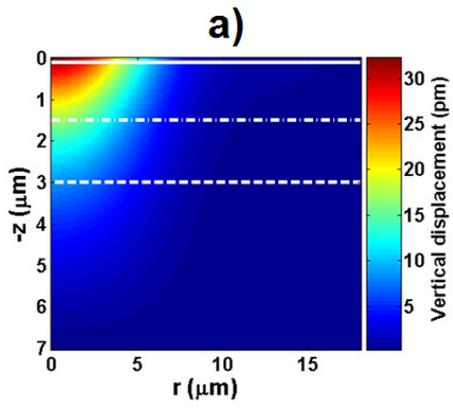

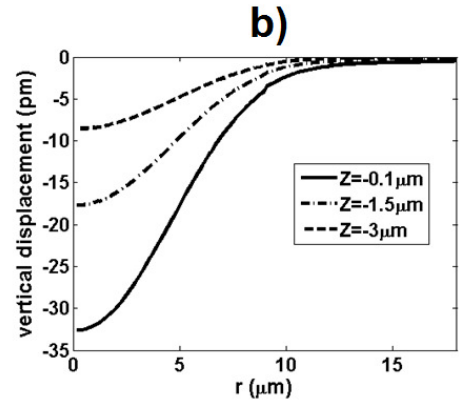

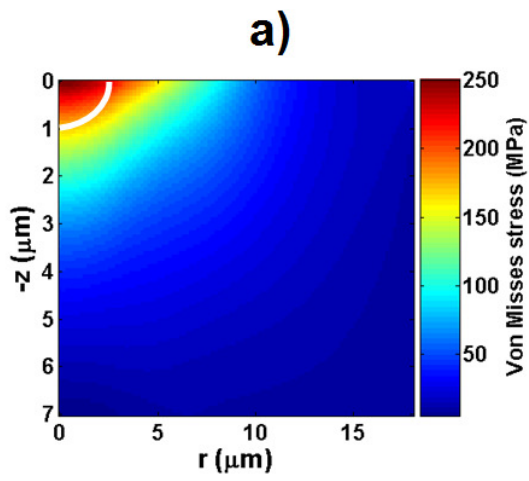

**a)**

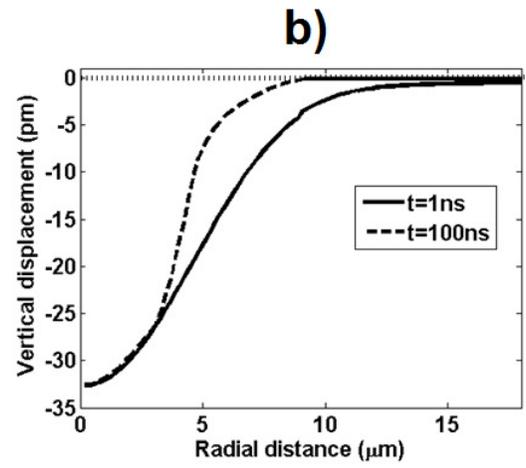

**b)**

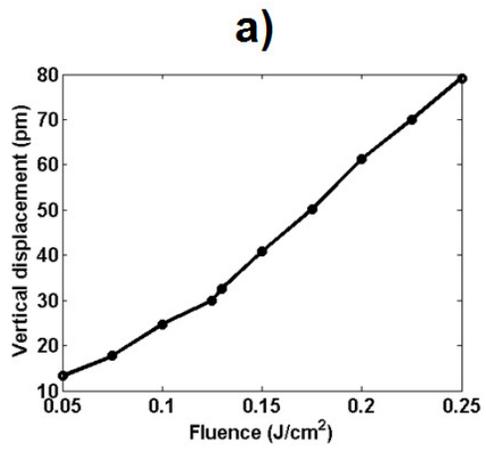

**a)**

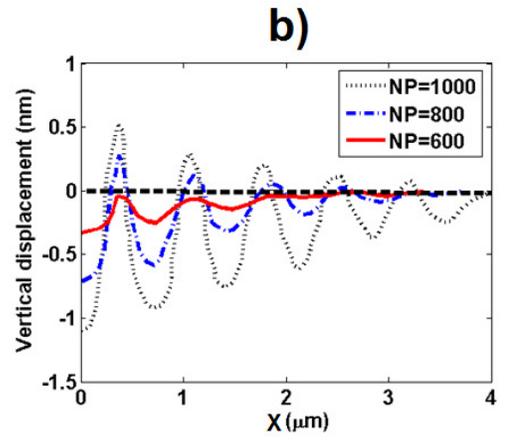

**b)**

**a)**

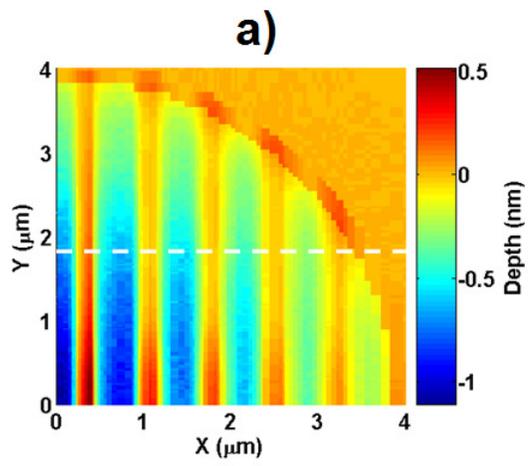

**b)**

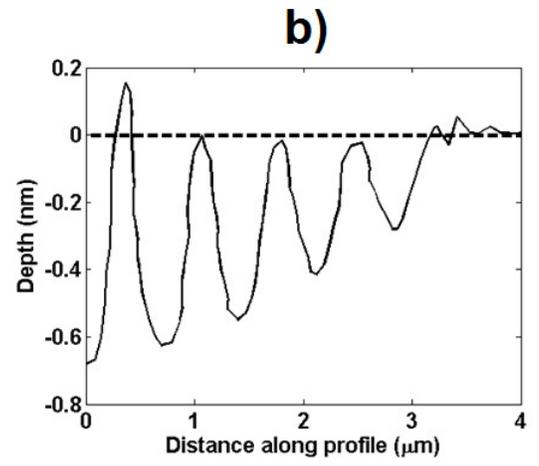

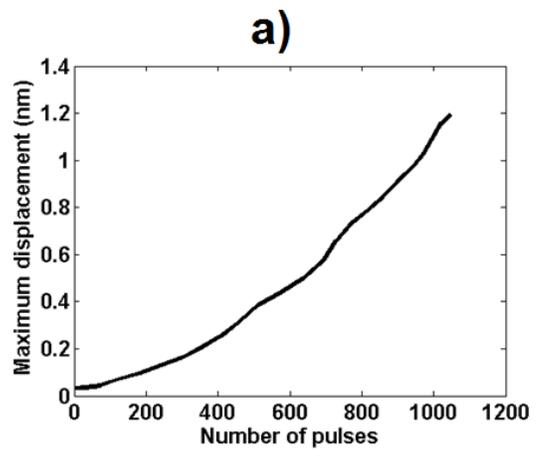

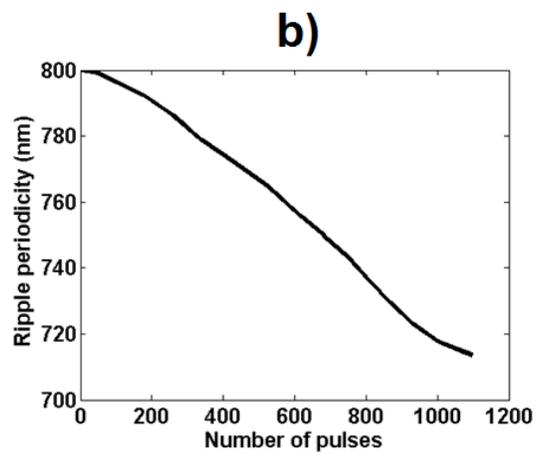

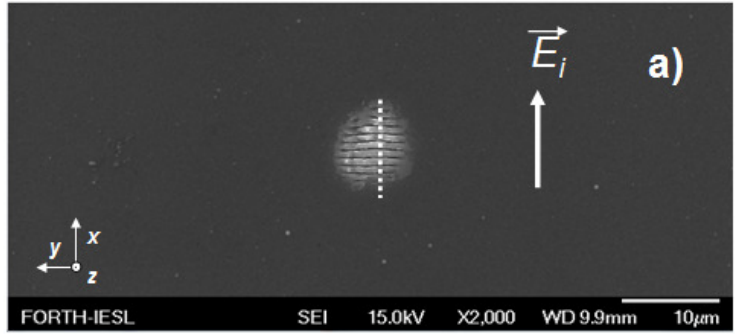

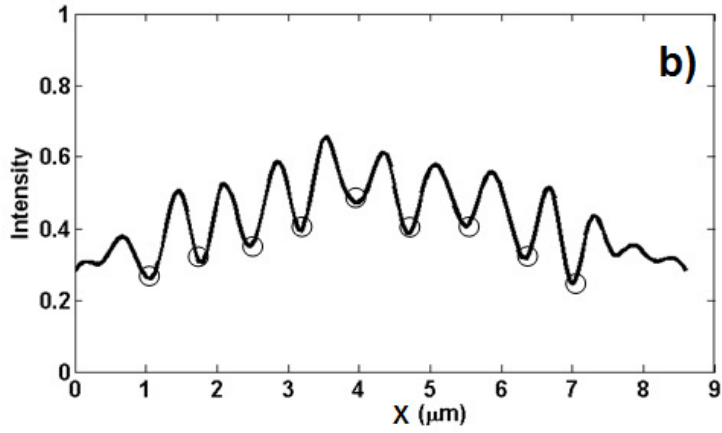

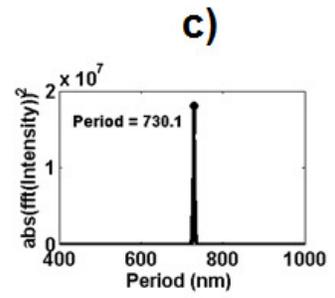

---

♣ Corresponding author, E-mail address: tsibidis@iesl.forth.gr